\documentclass[conference]{IEEEtran}
\IEEEoverridecommandlockouts
\usepackage{cite}
\usepackage{amsmath,amssymb,amsfonts}
\usepackage{algorithmic}
\usepackage{graphicx}
\usepackage{textcomp}
\usepackage{xcolor}
\usepackage{hyperref}
\usepackage{cuted}

\definecolor{linkred}{HTML}{a11616}
\definecolor{linkgreen}{HTML}{16a116}
\definecolor{linkblue}{HTML}{1616a1}
\hypersetup{
	colorlinks=true,
	linkcolor=linkred,
	filecolor=linkblue,
	urlcolor=linkblue,
	citecolor=linkgreen
}

\renewcommand{\vec}[1]{\boldsymbol{#1}}
\newcommand{\mat}[1]{\boldsymbol{#1}}
\renewcommand{\a}{\vec{a}}
\renewcommand{\b}{\vec{b}}
\renewcommand{\u}{\vec{u}}

\usepackage{adjustbox}

\usepackage{tikz, pgfplots}
\pgfplotsset{compat=newest}
\pgfplotsset{
    every axis/.append style={semithick},
    every axis plot/.append style={line width=1.2pt},
    every axis plot post/.append style={%
        every mark/.append style={line width=0.6pt},
    }
}

\usepackage{lipsum}

\def\BibTeX{{\rm B\kern-.05em{\sc i\kern-.025em b}\kern-.08em
		T\kern-.1667em\lower.7ex\hbox{E}\kern-.125emX}}
\begin{document}
	
	\title{Reviving Etter method for autoregressive inpainting: Generalization, evaluation, implementation%
    \thanks{The work was supported by the Czech Science Foundation (GAČR) Project No.\ 23-07294S. The authors want to thank all participants of the listening tests for their invaluable help.}
	}
	
	\author{%
		\IEEEauthorblockN{Ondřej Mokrý}
		\IEEEauthorblockA{\textit{Dept.\ of Telecommunications} \\
			\textit{Brno University of Technology}\\
			Czech Republic\\
			\href{mailto:ondrej.mokry@vut.cz}{ondrej.mokry@vut.cz}}
		\and
		\IEEEauthorblockN{Matěj Hrdlička}
		\IEEEauthorblockA{\textit{Dept.\ of Telecommunications} \\
			\textit{Brno University of Technology}\\
			Czech Republic\\
			\href{mailto:mates.hrd@gmail.com}{mates.hrd@gmail.com}}
		\and
		\IEEEauthorblockN{Pavel Rajmic}
		\IEEEauthorblockA{\textit{Dept.\ of Telecommunications} \\
			\textit{Brno University of Technology}\\
			Czech Republic\\
			\href{mailto:pavel.rajmic@vut.cz}{pavel.rajmic@vut.cz}}
	}
	
	\maketitle
	
	\begin{abstract}
        Audio inpainting aims to restore missing segments in an audio waveform, as encountered in dropouts and packet losses.
        This paper revisits the autoregressive (AR) interpolation method proposed by Etter, which combines forward and backward AR prediction through a structured linear system,
        yet lacks a widely used full implementation for audio signals.
        We provide an open implementation and propose two practical extensions: A~formulation that allows high-order AR models even when the gap length is shorter than the model order,
        and a~causal variant tailored to packet loss concealment.
        The method is evaluated on musical excerpts with gaps up to 80\,ms, using SNR and perceptually motivated objective difference grades,
        complemented by a listening test.
        The results show that Etter inpainting matches the performance of most AR- and sparsity-based baseline methods, with the exception of the iterative, gap-wise Janssen approach.
	\end{abstract}
	
	\begin{IEEEkeywords}
		audio,
        autoregression,
        implementation,
        inpainting,
        interpolation,
        packet loss concealment
	\end{IEEEkeywords}


\noindent
%

	\section{Introduction}
    \label{sec:introduction}

    Audio inpainting addresses the problem of filling (interpolating) missing segments of an audio signal.
    Such gaps arise in practical scenarios ranging from dropouts in field recordings to packet losses in networked transmission, where even short interruptions can cause clearly audible artifacts.
    Existing methods encompass techniques based on short-time Fourier spectrum sparsity \cite{Adler2012:Audio.inpainting, MokryRajmic2020:Inpainting.revisited}, including the so-called social sparsity variants \cite{kowalski2012social, Lieb2018:Audio.Inpainting}, dictionary-learning extensions~\cite{TaubockRajbamshiBalasz2021:SPAINMOD}, as well as approaches that exploit low-rank/factorization structure~\cite{Mokry2022:Audio.inpainting.NMF}.
    Among the oldest yet still prominent techniques are methods based on the autoregressive (AR) modelling, such as the Janssen iterative method \cite{javevr86, Oudre2018:Janssen.implementation, MokryRajmic2025:Inpainting.AR}. Besides their conceptual simplicity, AR methods remain attractive when low latency and modest computational cost are decisive, as in packet loss concealment (PLC), including recent hybrid approaches \cite{Mezza2024:Hybrid.packet.loss.concealment}.

    The present paper revisits the AR-based interpolation method of Etter \cite{Etter1996:Interpolation_AR}.
    Despite being widely cited, it has only rarely been applied in its full, non-simplified form, as in \cite{Clark2008:Modulation.decomposition.interpolation, Jiao2019:Atrial.activity.signal.reconstruction}.
    To the best of our knowledge, there is no publicly available implementation and no prior application to raw audio, which limits reproducible evaluation and wider adoption.

    The remainder of the paper is structured as follows.
    Sec.~\ref{sec:etter.overview} summarizes the original derivation of the Etter method, Sec.~\ref{sec:modifications} describes the proposed practical modifications, and Sec.~\ref{sec:experiments} reports the experimental evaluation.
    A faithful implementation of the full method is released in an external repository\footnote{\url{https://github.com/ondrejmokry/InpaintingAutoregressive}}\!.
    Throughout the paper, we use the following notation, stemming from the notation of the original paper \cite{Etter1996:Interpolation_AR}:
    Scalars are denoted in italics, matrices and vectors in bold with elements denoted using subscripts; signals are considered column vectors.
    Indexing starts from 1 with the only exception of the AR coefficient vectors $\a$ and $\b$ that include implicitly the element $a_0 = b_0 = -1$, following the convention of Etter~\cite{Etter1996:Interpolation_AR}.
    The input signal samples are $\vec{x} = [x_{1}, x_{2},\dots,x_{N}]^\top$\! with a~single, compact segment $\{x_{l},x_{l+1},\dots,x_{l+M-1}\}$ of length $M$ of missing samples.
    
    The AR model of order $K$ for the signal consists of the AR coefficients $a_1,\dots,a_K$, satisfying
    \begin{equation}
        x_{i} = \sum_{j=1}^{K} a_j x_{i-j} + u_{i}
        \label{eq:AR}
    \end{equation}
    where $\u$ is the excitation signal that includes the model error~\cite{Etter1996:Interpolation_AR}.
    Equivalently, the model may be viewed as an all-pole (IIR) synthesis mechanism (filter) driven by the excitation \cite[Ch.\,4]{AudioSignalProcessingAndCoding}.
    Optionally, $x_{i}$ in \eqref{eq:AR} can be included in the sum on the right-hand side by setting $a_0 = -1$.
    The classic estimate of $\a$ minimizes the residual energy 
    $\varepsilon = \u^\top\u$.
   
    \section{Overview of Etter interpolation method}
    \label{sec:etter.overview}

    This section summarizes the principal steps of the derivation of the Etter method from \cite[Sec.\,IV.]{Etter1996:Interpolation_AR}.

    \subsection{Estimating missing samples using left-sided and right-sided AR coefficients}
    \label{sec:signal.estimation}

    Denote $\hat{\vec{x}} = [\hat{x}_{l}, \hat{x}_{l+1}, \dots, \hat{x}_{l+M-1}]^\top$\!
    the column vector of samples to be interpolated,
    which are estimated using separate AR models for the left-sided and right-sided known signal portions.
    The left-sided AR coefficients $\a = [a_0, a_1, \dots, a_K]^\top$ are used to compute the forward prediction,
    \begin{equation}
        \hat{\vec{x}}^{\textup{L}} = [\hat{x}^{\textup{L}}_{l}, \hat{x}^{\textup{L}}_{l+1}, \dots, \hat{x}^{\textup{L}}_{l+M-1}]^\top\!,
    \end{equation}
    while the right-sided AR coefficients $\b = [b_0, b_1, \dots, b_K]^\top$ define the backward prediction,
    \begin{equation}
    	\hat{\vec{x}}^{\textup{R}} = [\hat{x}^{\textup{R}}_{l}, \hat{x}^{\textup{R}}_{l+1}, \dots, \hat{x}^{\textup{R}}_{l+M-1}]^\top\!.
    \end{equation}
    Note that estimating $\a$ and $\b$ themselves is addressed later in Sec.~\ref{sec:ar.estimation}.
    
    The corresponding residual (excitation) vectors
    $\hat{\u}^{\textup{L}}$ and $\hat{\u}^{\textup{R}}$
    over the missing segment are defined as:
    \begin{align}
    	\hat{\u}^{\textup{L}} &= [\hat{u}^{\textup{L}}_{l}, \hat{u}^{\textup{L}}_{l+1}, \dots, \hat{u}^{\textup{L}}_{l+M-1}]^\top\!,
        \label{eq:excitation:L}\\
    	\hat{\u}^{\textup{R}} &= [\hat{u}^{\textup{R}}_{l}, \hat{u}^{\textup{R}}_{l+1}, \dots, \hat{u}^{\textup{R}}_{l+M-1}]^\top\!.
        \label{eq:excitation:R}
    \end{align}
    
    To separate the contributions of the unknown samples in the missing gap from the surrounding known samples, filter matrices $\mat{A}$ and $\mat{B}$ and known-data matrices $\mat{L}$ and $\mat{R}$ are constructed below assuming $M > K$ (see Sec.~\ref{sec:modifications} for a~relaxation of this assumption).
    
    The forward filter matrix $\mat{A}$ is an $M \times M$ lower-triangular Toeplitz matrix formed from the forward AR parameters:
    \begin{equation}
        \label{eq:A}
    	\mat{A} = \begin{bmatrix}
    		a_0 & 0 & 0 & \dots & 0 \\
    		a_1 & a_0 & 0 & \dots & 0 \\
    		a_2 & a_1 & a_0 & \dots & 0 \\
    		\vdots & \vdots & \vdots & \ddots & \vdots \\
    		a_K & a_{K-1} & \dots & \dots & 0 \\
    		0 & a_K & a_{K-1} & \dots & 0 \\
    		\vdots & \vdots & \vdots & \ddots & \vdots \\
    		0 & 0 & \dots & a_1 & a_0
    	\end{bmatrix}.
    \end{equation}
    Likewise, the backward filter matrix $\mat{B}$ is an $M \times M$ upper-triangular matrix formed from the backward AR parameters:
    \begin{equation}
        \label{eq:B}
    	\mat{B} = \begin{bmatrix}
    		b_0 & b_1 & b_2 & \dots & b_K & \dots & 0 \\
    		0 & b_0 & b_1 & \dots & b_{K-1} & \dots & 0 \\
    		0 & 0 & b_0 & \dots & \dots & \dots & 0 \\
    		\vdots & \vdots & \vdots & \ddots & \vdots & \ddots & \vdots \\
    		0 & 0 & \dots & \dots & b_0 & b_1 & b_2 \\
    		0 & 0 & \dots & \dots & 0 & b_0 & b_1 \\
    		0 & 0 & \dots & \dots & 0 & 0 & b_0
    	\end{bmatrix}.
    \end{equation}
    
    The left known-data matrix $\mat{L}$ is an $M \times (K+1)$ matrix containing the known samples preceding the gap:
    \begin{equation}
    	\mat{L} = \begin{bmatrix}
    		0 & x_{l-1} & x_{l-2} & \dots & x_{l-K} \\
    		0 & 0 & x_{l-1} & \dots & x_{l-K+1} \\
    		0 & 0 & 0 & \dots & x_{l-K+2} \\
    		\vdots & \vdots & \vdots & \ddots & \vdots \\
    		0 & 0 & 0 & \dots & x_{l-1} \\
    		0 & 0 & 0 & \dots & 0 \\
    		\vdots & \vdots & \vdots & \ddots & \vdots \\
    		0 & 0 & 0 & \dots & 0
    	\end{bmatrix}.
    \end{equation}
    The right known-data matrix $\mat{R}$ is an $M \times (K+1)$ matrix containing the known samples succeeding the gap:
    \begin{equation}
    	\mat{R} = \begin{bmatrix}
    		0 & 0 & 0 & \dots & 0 \\
    		\vdots & \vdots & \vdots & \ddots & \vdots \\
    		0 & 0 & 0 & \dots & 0 \\
    		0 & 0 & \dots & \dots & x_{l+M} \\
    		\vdots & \vdots & \vdots & \ddots & \vdots \\
    		0 & 0 & x_{l+M} & \dots & x_{l+M+K-2} \\
    		0 & x_{l+M} & x_{l+M+1} & \dots & x_{l+M+K-1}
    	\end{bmatrix}.
    \end{equation}
    
    Using these definitions,
    the residuals \eqref{eq:excitation:L} and \eqref{eq:excitation:R} can be decomposed into the contribution of the known samples and the estimated samples as
    \begin{align}
        \hat{\u}^{\textup{L}} = -\mat{L}\a - \mat{A}\hat{\vec{x}}^{\textup{L}},
        \label{eq:excitation.mat:L}\\
        \hat{\u}^{\textup{R}} = -\mat{R}\b - \mat{B}\hat{\vec{x}}^{\textup{R}}.
        \label{eq:excitation.mat:R}
    \end{align}
    
    To achieve joint estimation across the gap, a unified target vector $\hat{\vec{x}} = \hat{\vec{x}}^{\textup{L}} = \hat{\vec{x}}^{\textup{R}}$ is required, such that it minimizes the total residual energy (cf.\ \eqref{eq:AR} where we search for the AR coefficients)
    \begin{equation}
    	\varepsilon = (\hat{\u}^{\textup{L}})^\top \hat{\u}^{\textup{L}} + (\hat{\u}^{\textup{R}})^\top \hat{\u}^{\textup{R}}.
    \end{equation}
    It can be shown that such a~vector is the solution to the linear system
    \begin{equation}
    	\label{eq:Dxy}
    	\mat{D} \hat{\vec{x}} = \vec{y}
    \end{equation}
    where the symmetric matrix $\mat{D}$ is formed using \eqref{eq:A} and \eqref{eq:B} as
    \begin{equation}
    	\label{eq:D}
    	\mat{D} = \mat{A}^\top \mat{A} + \mat{B}^\top \mat{B}
    \end{equation}
    and the vector $\vec{y}$ incorporates the boundary conditions from both the left and right known signal segments:
    \begin{equation}
    	\label{eq:y}
    	\vec{y} = -\mat{A}^\top \mat{L} \a - \mat{B}^\top \mat{R} \b.
    \end{equation}

    \subsection{Estimating AR model}
    \label{sec:ar.estimation}

    Although \cite{Etter1996:Interpolation_AR} does not specify a procedure how to estimate the left-sided and right-sided AR coefficients $\a$ and $\b$, two standard strategies can be employed.
    
    The first strategy minimizes the energy of the residual signal $\u$ in \eqref{eq:AR}.
    This yields a~convex optimization problem that can be efficiently solved via the Levinson--Durbin algorithm~\cite{Durbin1960,Levinson1946}.
    This approach is typically referred to simply as the LPC algorithm,
    with LPC denoting linear prediction coefficients.
    
    In contrast to LPC, the Burg algorithm
    \cite{Burg1975:PHD.Maximum.entropy.spectral}
    for AR parameter estimation makes the additional assumption that the \emph{same} parameters should model
    both the signal $\vec{x}$ and its time-reversed counterpart,
    which ensures stability of the underlying all-pole filter
    \cite[Sec.\,12.3.3]{Proakis1996:DSP}.

    \subsection{Weighted extrapolation}
    \label{sec:extrapolation}

    The method described in Sec.\ \ref{sec:signal.estimation} requires a~matrix inversion when solving for the missing-sample estimate $\hat{\vec{x}}$.
    As a~computationally cheaper alternative, Etter proposes a~suboptimal estimate by cross-fading the \emph{independent} forward and backward predictors $\hat{\vec{x}}^{\textup{L}}$ and $\hat{\vec{x}}^{\textup{R}}$ using suitable weights \cite[Sec.\,IV.]{Etter1996:Interpolation_AR}.
    In this construction, the excitation vectors $\hat{\u}^{\textup{L}}$ and $\hat{\u}^{\textup{R}}$ over the gap are neglected ($\hat{\u}^{\textup{L}}=\hat{\u}^{\textup{R}}=\vec{0}$).
    We refer to this approach as \emph{extrapolation-based inpainting}, or shortly \emph{extrapolation}.

    \section{Proposed modifications}
    \label{sec:modifications}
    
    This section presents modifications of the original Etter procedure aimed at improving its practical applicability.
    In particular, Sec.~\ref{sec:modifications:MK} generalizes the derivation to cover the case $M\leq K$ (i.e., the AR model order exceeding the gap length)
    and Sec.~\ref{sec:modifications:plc} discusses further adjustments motivated by typical PLC constraints.

    \subsection{Generalization for $M\leq K$}
    \label{sec:modifications:MK}
	In the case $M\leq K$, the construction of the filter and data matrices in Sec.~\ref{sec:etter.overview} can be retained by appropriately truncating the Toeplitz structures to the gap length.
    First, the lower-triangular forward prediction filter matrix is defined as
	\begin{subequations}		
    \label{eq:matrix_A_system}
		\begin{equation}
			\label{eq:matrix_A}
			\mat{A} = \begin{bmatrix}
				a_0 & 0 & 0 & \dots & 0 \\
				a_1 & a_0 & 0 & \dots & 0 \\
				a_2 & a_1 & a_0 & \dots & 0 \\
				\vdots & \vdots & \vdots & \ddots & \vdots \\
				a_{M-1} & a_{M-2} & a_{M-3} & \dots & a_0
			\end{bmatrix},
		\end{equation}
	    or, in an explicit element-wise form,
		\begin{equation}
			\label{eq:elem_A}
			A_{i,j} = \begin{cases} 
				a_{i-j}, & \text{if } i \ge j \\ 
				0, & \text{if } i < j 
			\end{cases}
		\end{equation}
		for $i, j \in \{1, \dots, M\}$.
	\end{subequations}
	The backward prediction filter is represented by the upper-triangular matrix
	\begin{subequations}		
    \label{eq:matrix_B_system}
		\begin{equation}
			\label{eq:matrix_B}
			\mat{B} = \begin{bmatrix}
				b_0 & b_1 & b_2 & \dots & b_{M-1} \\
				0 & b_0 & b_1 & \dots & b_{M-2} \\
				0 & 0 & b_0 & \dots & b_{M-3} \\
				\vdots & \vdots & \vdots & \ddots & \vdots \\
				0 & 0 & 0 & \dots & b_0
			\end{bmatrix},
		\end{equation}
        i.e.,
		\begin{equation}
			\label{eq:elem_B}
			B_{i,j} = \begin{cases} 
				b_{j-i}, & \text{if } j \ge i \\ 
				0, & \text{if } j < i 
			\end{cases}
		\end{equation}
		for $i, j \in \{1, \dots, M\}$.
	\end{subequations}
		
	The surrounding known samples form the left- and right-context matrices $\mat{L}$ and $\mat{R}$.
    In particular,
    for the left context, it holds
		\begin{subequations}			
        \label{eq:matrix_L_system}
			{\setlength{\arraycolsep}{3pt}%
			\begin{equation}
				\label{eq:matrix_L}
				\mat{L} = \begin{bmatrix}
					0 & x_{l-1} & x_{l-2} & \dots & x_{l-M} & \dots & x_{l-K} \\
					0 & 0 & x_{l-1} & \dots & x_{l-M+1} & \dots & x_{l-K+1} \\
					0 & 0 & 0 & \dots & x_{l-M+2} & \dots & x_{l-K+2} \\
					\vdots & \vdots & \vdots & \ddots & \vdots & \ddots & \vdots \\
					0 & 0 & 0 & \dots & x_{l-1} & \dots & x_{l-K+M-1}
				\end{bmatrix},
			\end{equation}
			}%
			\begin{equation}
				\label{eq:elem_L}
				L_{i,j} = \begin{cases} 
					0, & \text{if } j \le i \\ 
					x_{l + i - j}, & \text{if } j > i 
				\end{cases}
			\end{equation}
			for $i \in \{1, \dots, M\}$ and $j \in \{1, \dots, K+1\}$.
		\end{subequations}
		The contribution of the unknown samples to the forward prediction can be written using the lower-triangular matrix
		\begin{subequations}
			\label{eq:matrix_XL_hat_gen_system}
			\begin{equation}
				\label{eq:matrix_XL_hat_gen}
				\hat{\mat{X}}^{\textup{L}} = 
				\begin{bmatrix}
					\hat{x}^{\textup{L}}_{l} & 0 & \dots & 0 & 0 & \dots & 0 \\
					\hat{x}^{\textup{L}}_{l+1} & \hat{x}^{\textup{L}}_{l} & \dots & 0 & 0 & \dots & 0 \\
					\vdots & \vdots & \ddots & \vdots & \vdots & \ddots & \vdots \\
					\hat{x}^{\textup{L}}_{l+M-1} & \hat{x}^{\textup{L}}_{l+M-2} & \dots & \hat{x}^{\textup{L}}_{l} & 0 & \dots & 0
				\end{bmatrix},
			\end{equation}
			\begin{equation}
				\label{eq:elem_XL_hat_gen}
				(\hat{X}^{\textup{L}})_{i,j} = \begin{cases} 
					\hat{x}^{\textup{L}}_{l + i - j}, & \text{if } j \le i \\ 
					0, & \text{if } j > i 
				\end{cases}
			\end{equation}
			for $i \in \{1, \dots, M\}$ and $j \in \{1, \dots, K+1\}$.
		\end{subequations}
		Analogously, the right-sided known context is embedded into the matrix
		\begin{subequations}
			\label{eq:matrix_R_system}
			{\setlength{\arraycolsep}{3pt}%
			\begin{equation}
				\label{eq:matrix_R}
				\mat{R} = 
                \begin{bmatrix}
					0 & \dots & 0 & x_{l+M} & x_{l+M+1} & \dots & x_{l+K} \\
					0 & \dots & x_{l+M} & x_{l+M+1} & x_{l+M+2} & \dots & x_{l+K+1} \\
					\vdots & \ddots & \vdots & \vdots & \vdots & \ddots & \vdots \\
					0 & x_{l+M} & \dots & \dots & \dots & \dots & x_{l+M+K-1}
				\end{bmatrix},
			\end{equation}
			}%
			\begin{equation}
				\label{eq:elem_R}
				R_{i,j} = \begin{cases} 
					0, & \text{if } j \le M - i + 1 \\ 
					x_{l + i + j - 2}, & \text{if } j > M - i + 1 
				\end{cases}
			\end{equation}
			for $i \in \{1, \dots, M\}$ and $j \in \{1, \dots, K+1\}$.
		\end{subequations}
		Finally, the unknown samples enter the backward prediction through the structured matrix
		\begin{subequations}
			\label{eq:matrix_XR_hat_gen_system}
			{\setlength{\arraycolsep}{3pt}%
			\begin{equation}
				\label{eq:matrix_XR_hat_gen}
				\hat{\mat{X}}^{\textup{R}} =
				\begin{bmatrix}
					\hat{x}^{\textup{R}}_{l} & \hat{x}^{\textup{R}}_{l+1} & \dots & \hat{x}^{\textup{R}}_{l+M-1} & 0 & \dots & 0 \\
					\hat{x}^{\textup{R}}_{l+1} & \hat{x}^{\textup{R}}_{l+2} & \dots & 0 & 0 & \dots & 0 \\
					\vdots & \vdots & \ddots & \vdots & \vdots & \ddots & \vdots \\
					\hat{x}^{\textup{R}}_{l+M-1} & 0 & \dots & 0 & 0 & \dots & 0
				\end{bmatrix},
			\end{equation}
			}%
			\begin{equation}
				\label{eq:elem_XR_hat_gen}
				(\hat{X}^{\textup{R}})_{i,j} = \begin{cases} 
					\hat{x}^{\textup{R}}_{l + i + j - 2}, & \text{if } j \le M - i + 1 \\ 
					0, & \text{if } j > M - i + 1 
				\end{cases}
			\end{equation}
			for $i \in \{1, \dots, M\}$ and $j \in \{1, \dots, K+1\}$.
		\end{subequations}
    Although the matrices slightly differ from those in their original formulation, the interpolated samples are still computed by solving the same linear system \eqref{eq:Dxy} as in Sec.~\ref{sec:etter.overview}.

    \subsection{PLC case}
    \label{sec:modifications:plc}
    
    In low-latency PLC setting, it is often necessary to rely only on the left-sided context, without using a right-sided AR model.
    In such a~configuration, the estimate of the missing segment follows directly from \eqref{eq:excitation.mat:L} as
    \begin{equation}
        \label{eq:plc:left.only}
        \hat{\vec{x}} = \hat{\vec{x}}^{\textup{L}} = -\mat{A}^{-1}\mat{L}\a.
    \end{equation}
    This expression is equivalent to simple forward extrapolation using the left-sided AR parameters: one predicts the samples sequentially while omitting the excitation term in \eqref{eq:AR}.
    Note that \eqref{eq:plc:left.only} also formalizes the forward part of the extrapolation-based method described in Sec.\ \ref{sec:extrapolation}.
    
    In practical PLC, a smooth transition to the first available sample \textit{after} the gap is important to avoid click-like artifacts; this can be achieved, for example, via short crossfades as in related audio restoration problems such as declipping \cite{ZaviskaRajmicMokry2022:Declipping.crossfading}.

    \section{Experiments and results}
    \label{sec:experiments}
    
    The experiments adhere to the protocol of \cite{MokryRajmic2025:Inpainting.AR}.
    We use a~subset of EBU SQAM\footnote{\url{https://github.com/ondrejmokry/TestSignals}} \cite{EBUSQAM},
    and estimate the AR coefficients from a~context of 4096 samples on each side of a~gap.
    To objectively quantify reconstruction quality, we compare each reconstructed signal $\hat{\vec{x}}$ with the respective clean reference $\vec{x}^{\textup{ref}}$ using standard measures.
    The signal-to-noise ratio is computed in decibels as \cite{Adler2012:Audio.inpainting}
    \begin{equation}
        \text{SNR}(\hat{\vec{x}},\vec{x}^{\textup{ref}})=10\log_{10}\!\frac{\lVert\vec{x}^{\textup{ref}}\rVert_2^2}{\lVert\vec{x}^{\textup{ref}}-\hat{\vec{x}}\rVert_2^2}.
    \end{equation}
    To incorporate psychoacoustic effects beyond sample-wise similarity,
    we use the objective difference grade (ODG) obtained with either PEMO-Q \cite{Huber:2006a} or PEAQ \cite{Kabal2002:PEAQ}, predicting perceived degradation on a~scale from $-4$ (very annoying) to $0$ (imperceptible).

\begin{figure}[htb]
        \centering
        \adjustbox{width=\linewidth}{%
        \input{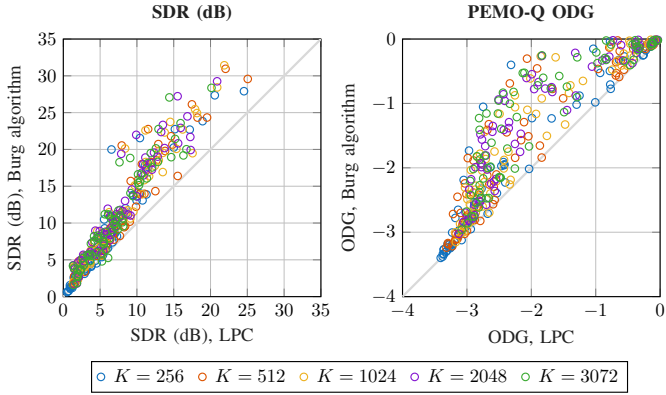}}
        \caption{Evaluating the AR model estimators.}
        \label{fig:maintest_scatter_shorter}
    \end{figure}

    First, we compare the two AR model estimators from Sec.~\ref{sec:ar.estimation}: LPC (implemented via the Matlab function \texttt{lpc}\footnote{\url{https://www.mathworks.com/help/signal/ref/lpc.html}}) and the Burg algorithm (\texttt{arburg}\footnote{\url{https://www.mathworks.com/help/signal/ref/arburg.html}}).
    As shown in Fig.~\ref{fig:maintest_scatter_shorter}, the Burg algorithm consistently achieves higher reconstruction quality than LPC. This is supported by a~Wilcoxon signed-rank test \cite{Hollander1999}, which rejects equality of medians at the 5\% significance level in favor of Burg in all cases.

\begin{figure}[htb]
        \centering
        \adjustbox{width=\linewidth}{%
%
%
\definecolor{mycolor1}{rgb}{0.62640,0.77720,0.89800}%
\definecolor{mycolor2}{rgb}{0.06600,0.44300,0.74500}%
\definecolor{mycolor3}{rgb}{0.94640,0.73160,0.60000}%
\definecolor{mycolor4}{rgb}{0.86600,0.32900,0.00000}%
\definecolor{mycolor5}{rgb}{0.97160,0.87760,0.65000}%
\definecolor{mycolor6}{rgb}{0.92900,0.69400,0.12500}%
\definecolor{mycolor7}{rgb}{0.80840,0.63440,0.92760}%
\definecolor{mycolor8}{rgb}{0.52100,0.08600,0.81900}%
\definecolor{mycolor9}{rgb}{0.69240,0.86640,0.67840}%
\definecolor{mycolor10}{rgb}{0.23100,0.66600,0.19600}%
\definecolor{mycolor11}{rgb}{0.12941,0.12941,0.12941}%
\begin{tikzpicture}
	
	\begin{axis}[%
		width=1.90in,
		height=1.90in,
		at={(0.748in,1.357in)},
		scale only axis,
		xmin=10,
		xmax=80,
		xlabel style={font=\color{mycolor11}},
		xlabel={gap length (ms)},
		ymin=0,
		ymax=20,
		ylabel style={font=\color{mycolor11}},
		ylabel={SDR (dB)},
		axis background/.style={fill=white},
		title style={font=\bfseries\color{mycolor11}},
		title={SDR (dB)},
		xmajorgrids,
		ymajorgrids,
		legend style={
            at={(1.15,-0.25)},
            anchor=north,
            legend cell align=left,
            align=left,
            legend columns=5,
            /tikz/every even column/.append style={column sep=2mm},
            /tikz/every odd column/.append style={column sep=1mm}}
		]
		\addplot [color=mycolor1]
		table[row sep=crcr]{%
			10	14.4622890775826\\
			20	11.1997157478168\\
			30	9.39649370526907\\
			40	8.00122259348286\\
			50	6.75495555230516\\
			60	5.45083590580567\\
			70	4.87261789305641\\
			80	4.67571532304612\\
		};
		\addlegendentry{$K = 256$}
		
		\addplot [color=mycolor2, dashed, forget plot]
		table[row sep=crcr]{%
			10	11.3863623291256\\
			20	8.57559700957565\\
			30	6.50745284113094\\
			40	5.08101622566743\\
			50	4.95930716123804\\
			60	4.14863482776828\\
			70	3.54633762755456\\
			80	3.3909037505112\\
		};
		\addplot [color=mycolor3]
		table[row sep=crcr]{%
			10	17.8915521122505\\
			20	13.9376612797936\\
			30	12.1432614416303\\
			40	10.5196186065876\\
			50	8.72619418343476\\
			60	7.17092790514808\\
			70	6.47255138416716\\
			80	6.10522657042069\\
		};
		\addlegendentry{$K = 512$}
		
		\addplot [color=mycolor4, dashed, forget plot]
		table[row sep=crcr]{%
			10	13.7216711958728\\
			20	10.5610286033636\\
			30	8.10479634863176\\
			40	6.57475220848332\\
			50	6.18475587671213\\
			60	5.29652027661564\\
			70	4.54089161739862\\
			80	4.35499184804817\\
		};
		\addplot [color=mycolor5]
		table[row sep=crcr]{%
			10	18.6793861059496\\
			20	14.9682314890386\\
			30	13.3667736208095\\
			40	11.6480782132041\\
			50	9.71995934642077\\
			60	8.1345185440278\\
			70	7.68025165382962\\
			80	7.09853340887298\\
		};
		\addlegendentry{$K = 1024$}
		
		\addplot [color=mycolor6, dashed, forget plot]
		table[row sep=crcr]{%
			10	13.7100158777799\\
			20	10.6648998410096\\
			30	8.44322976088703\\
			40	7.28920327130117\\
			50	6.72065480313209\\
			60	5.69060382848254\\
			70	5.01803810281495\\
			80	4.84165615994942\\
		};
		\addplot [color=mycolor7]
		table[row sep=crcr]{%
			10	18.2805000016224\\
			20	14.7950484199402\\
			30	13.5529731183995\\
			40	11.7496099861318\\
			50	10.1685040305065\\
			60	8.47971902138933\\
			70	7.95911348135684\\
			80	7.61357966755652\\
		};
		\addlegendentry{$K = 2048$}
		
		\addplot [color=mycolor8, dashed, forget plot]
		table[row sep=crcr]{%
			10	12.8046742638365\\
			20	10.2666432462554\\
			30	8.07609526778816\\
			40	7.21493653803776\\
			50	6.78016799792818\\
			60	5.75294403445581\\
			70	4.9863922157966\\
			80	4.80686108575313\\
		};
		\addplot [color=mycolor9]
		table[row sep=crcr]{%
			10	17.414895386394\\
			20	13.5474104912847\\
			30	12.5445500052214\\
			40	10.7627399890737\\
			50	9.48614863555468\\
			60	7.93062749649564\\
			70	7.46563515431299\\
			80	7.23759352043702\\
		};
		\addlegendentry{$K = 3072$}
		
		\addplot [color=mycolor10, dashed, forget plot]
		table[row sep=crcr]{%
			10	11.9979158347538\\
			20	9.7565495503059\\
			30	7.61124575072491\\
			40	6.85227909485104\\
			50	6.4470686396876\\
			60	5.44841995494873\\
			70	4.7768604991838\\
			80	4.56389672443893\\
		};
	\end{axis}

	\begin{axis}[%
		width=1.90in,
		height=1.90in,
		at={(3.25in,1.357in)},
		scale only axis,
		xmin=10,
		xmax=80,
		xlabel style={font=\color{mycolor11}},
		xlabel={gap length (ms)},
		ymin=-4,
		ymax=0,
		ylabel style={font=\color{mycolor11}},
		ylabel={ODG},
		axis background/.style={fill=white},
		title style={font=\bfseries\color{mycolor11}},
		title={PEMO-Q ODG},
		xmajorgrids,
		ymajorgrids,
		legend style={at={(0.03,0.03)}, anchor=south west, legend cell align=left, align=left}
		]
		\addplot [color=mycolor1]
		table[row sep=crcr]{%
			10	-0.111235261925889\\
			20	-0.518910088315483\\
			30	-1.40473578339427\\
			40	-2.16939280863131\\
			50	-2.60265013675721\\
			60	-2.86556566249858\\
			70	-3.04714598237061\\
			80	-3.12039692230151\\
		};
		
		\addplot [color=mycolor2, dashed, forget plot]
		table[row sep=crcr]{%
			10	-0.230774133553991\\
			20	-0.770851869823285\\
			30	-1.8666086074433\\
			40	-2.61212982414544\\
			50	-2.95007536451616\\
			60	-3.09397249546528\\
			70	-3.18057574637772\\
			80	-3.22919710326495\\
		};
		\addplot [color=mycolor3]
		table[row sep=crcr]{%
			10	-0.0621166384343602\\
			20	-0.313402651165602\\
			30	-0.943603672064947\\
			40	-1.60455632088663\\
			50	-2.18585471178034\\
			60	-2.52686084998343\\
			70	-2.76969584682051\\
			80	-2.94734966692709\\
		};
		
		\addplot [color=mycolor4, dashed, forget plot]
		table[row sep=crcr]{%
			10	-0.172936872294599\\
			20	-0.534052468559641\\
			30	-1.5716838866546\\
			40	-2.30942785691403\\
			50	-2.70812803482988\\
			60	-2.929069723925\\
			70	-3.06634130546937\\
			80	-3.14691737703032\\
		};
		\addplot [color=mycolor5]
		table[row sep=crcr]{%
			10	-0.0443274284058819\\
			20	-0.20502326973383\\
			30	-0.69265506149589\\
			40	-1.27523957496233\\
			50	-1.87609493949049\\
			60	-2.26725198807834\\
			70	-2.5248286794383\\
			80	-2.71373765541936\\
		};
		
		\addplot [color=mycolor6, dashed, forget plot]
		table[row sep=crcr]{%
			10	-0.14298908970885\\
			20	-0.448078495453974\\
			30	-1.31213481630648\\
			40	-2.02837796842674\\
			50	-2.53449059219083\\
			60	-2.79963445146581\\
			70	-2.94106111662405\\
			80	-3.01659444330136\\
		};
		\addplot [color=mycolor7]
		table[row sep=crcr]{%
			10	-0.0378933983314482\\
			20	-0.165856752911012\\
			30	-0.481664713075842\\
			40	-0.947541867528085\\
			50	-1.47042581400567\\
			60	-1.93319652016359\\
			70	-2.20915983462033\\
			80	-2.47918935818701\\
		};
		
		\addplot [color=mycolor8, dashed, forget plot]
		table[row sep=crcr]{%
			10	-0.187504286573906\\
			20	-0.394114884506911\\
			30	-1.1649832381564\\
			40	-1.91796017207849\\
			50	-2.40837720992646\\
			60	-2.65811043959446\\
			70	-2.84423241019842\\
			80	-2.92373828909958\\
		};
		\addplot [color=mycolor9]
		table[row sep=crcr]{%
			10	-0.0473021329530994\\
			20	-0.188202022015279\\
			30	-0.491977161393797\\
			40	-0.915310882664807\\
			50	-1.471010429003\\
			60	-1.74318274745913\\
			70	-2.18447412692954\\
			80	-2.35945554192997\\
		};
		
		\addplot [color=mycolor10, dashed, forget plot]
		table[row sep=crcr]{%
			10	-0.213504185943914\\
			20	-0.420787731449991\\
			30	-1.15330616391591\\
			40	-1.90690462931543\\
			50	-2.3721708266237\\
			60	-2.63797893321477\\
			70	-2.80940901237628\\
			80	-2.89141594448761\\
		};
	\end{axis}

\end{tikzpicture}%
}
        \caption{Testing the AR model order. Solid line represents results obtained using Burg algorithm as AR model estimator; dashed line represents LPC.}
        \label{fig:maintest_plot_by_method_shorter}
    \end{figure}
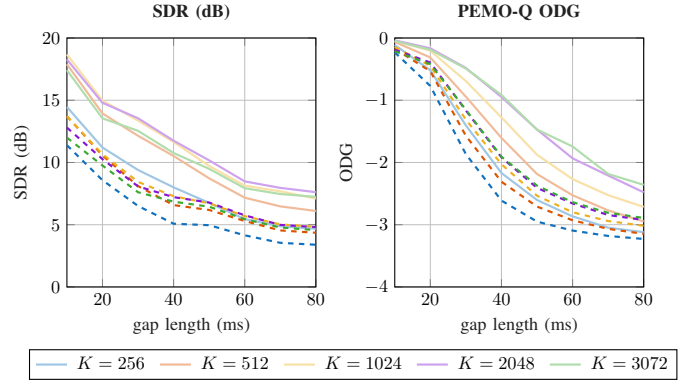

    As seen in Fig.~\ref{fig:maintest_plot_by_method_shorter}, the influence of the model order is similar to that observed for other AR methods \cite{MokryRajmic2025:Inpainting.AR}:
    the optimal order tends to be relatively high, but not the maximum tested.
    
\begin{figure}[htb]
        \centering
        \adjustbox{width=\linewidth}{%
        \input{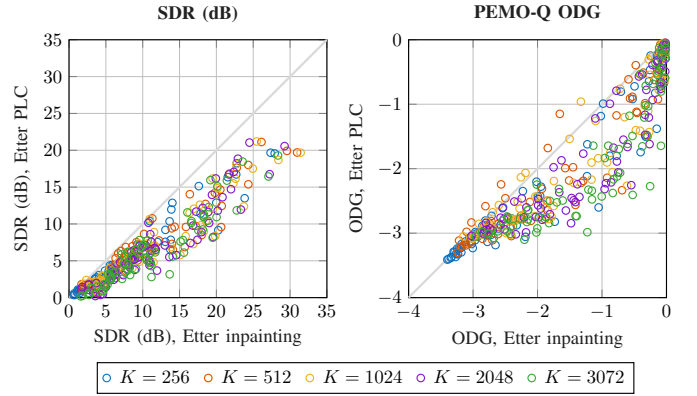}}
        \caption{Testing full (inpainting) and PLC mode of Etter method.}
        \label{fig:maintest_scatter_etter_shorter_plc}
    \end{figure}

\begin{figure}[htb]
        \centering
        \adjustbox{width=\linewidth}{%
        \input{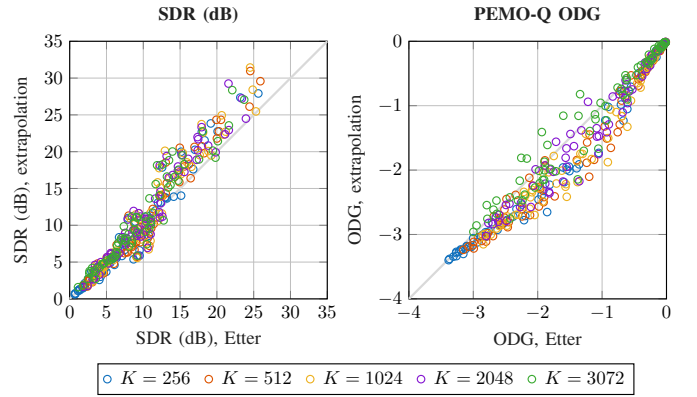}}
        \caption{Testing full and simplified (extrapolation) version of Etter method.}
        \label{fig:maintest_scatter_etter_shorter_extrapolation}
    \end{figure}

    Next, we compare the Etter method with its two simplified versions: the extrapolation method described in Sec.~\ref{sec:extrapolation}
    and the (causal) PLC flavor developed in Sec.~\ref{sec:modifications:plc}.
    Fig.~\ref{fig:maintest_scatter_etter_shorter_plc} indicates that the full Etter method (referred to as \emph{Etter inpainting}) outperforms the PLC variant.
    This is expected as the inpainting case considers the right context which can prove significant especially when non-stationarity appears in the region of the gap.
    On the other hand, the comparison of Etter with the simplified case of weighted extrapolation is intriguing,
    since Etter considers the extrapolation case as suboptimal \cite[Sec.\,IV.]{Etter1996:Interpolation_AR}.
    According to \mbox{PEMO-Q}, Etter inpainting performs better than extrapolation, except at $p=3072$ (see Fig.~\ref{fig:maintest_scatter_etter_shorter_extrapolation}).
    The remaining objective metrics suggest that extrapolation can be preferable at higher model orders, whereas at lower orders the results are inconclusive.


    \begin{figure*}
        \centering
        \adjustbox{width=\linewidth}{%
        \input{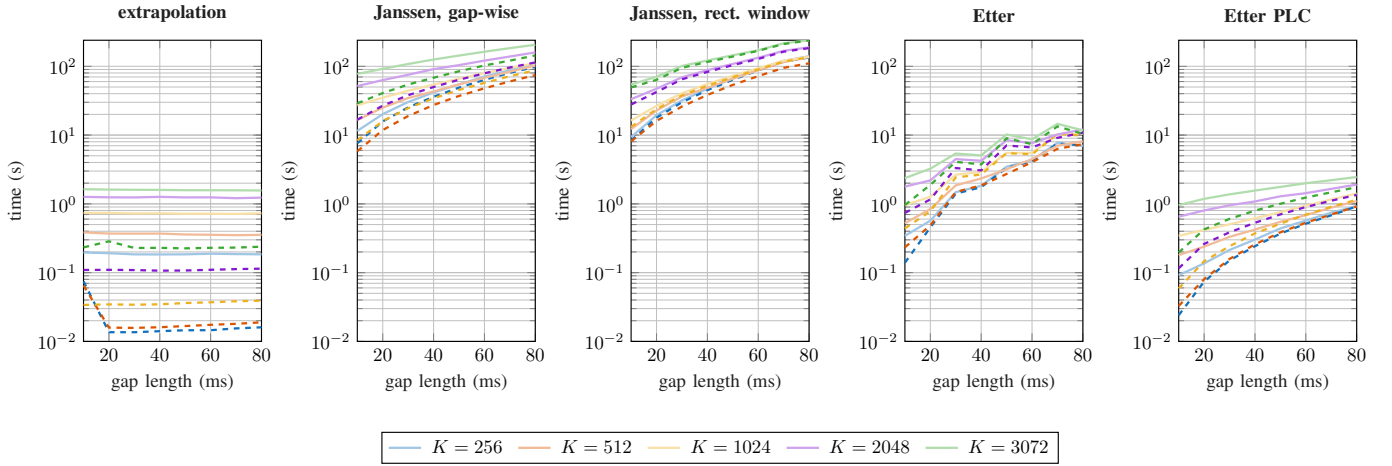}}
        \caption{Elapsed times of Etter and other AR-based methods depending on AR model order. As in Fig.~\ref{fig:maintest_plot_by_method_shorter}, solid and dashed lines represent AR model estimation with Burg algorithm and LPC, respectively.}
        \label{fig:maintest_plot_by_method_shorter_time}
    \end{figure*}

    \begin{figure*}
        \centering
        \adjustbox{width=\linewidth}{%
%
%
\definecolor{mycolor1}{rgb}{0.06600,0.44300,0.74500}%
\definecolor{mycolor2}{rgb}{0.86600,0.32900,0.00000}%
\definecolor{mycolor3}{rgb}{0.92900,0.69400,0.12500}%
\definecolor{mycolor4}{rgb}{0.52100,0.08600,0.81900}%
\definecolor{mycolor5}{rgb}{0.23100,0.66600,0.19600}%
\definecolor{mycolor6}{rgb}{0.18400,0.74500,0.93700}%
\definecolor{mycolor7}{rgb}{0.12941,0.12941,0.12941}%
\begin{tikzpicture}
	
	\begin{axis}[%
		width=2.80in,
		height=2.40in,
		at={(0in,0.469in)},
		scale only axis,
		xmin=10,
		xmax=80,
		xlabel style={font=\color{mycolor7}},
		xlabel={gap length (ms)},
		ymin=0,
		ymax=20,
		ylabel style={font=\color{mycolor7}},
		ylabel={SDR (dB)},
		axis background/.style={fill=white},
		title style={font=\bfseries\color{mycolor7}},
		title={SDR (dB)},
		xmajorgrids,
		ymajorgrids,
		legend style={legend cell align=left, align=left}
		]
		\addplot [color=mycolor1]
		table[row sep=crcr]{%
			10	16.1512418874303\\
			20	13.2864987910056\\
			30	11.9000268478401\\
			40	10.2082906132531\\
			50	9.42930248953038\\
			60	8.29283103648075\\
			70	7.83850468988678\\
			80	7.55247493059489\\
		};
		
		\addplot [color=mycolor2]
		table[row sep=crcr]{%
			10	18.6259135872454\\
			20	15.3457042779823\\
			30	13.7716787910352\\
			40	12.9128228990526\\
			50	11.4221370877055\\
			60	10.2685157361508\\
			70	9.75943177362145\\
			80	9.03881200155933\\
		};
		
		\addplot [color=mycolor3]
		table[row sep=crcr]{%
			10	18.8804568785158\\
            20	14.8006274084239\\
            30	12.6594993911984\\
            40	11.0056385293903\\
            50	7.31414007265942\\
            60	5.22950933717408\\
            70	4.42809967891603\\
            80	3.93545329253067\\
		};
		
		\addplot [color=mycolor4]
		table[row sep=crcr]{%
            10	18.2805000016224\\
            20	14.7950484199402\\
            30	13.5529731183995\\
            40	11.7496099861318\\
            50	10.1685040305065\\
            60	8.47971902138933\\
            70	7.95911348135684\\
            80	7.61357966755652\\
		};
		
		\addplot [color=mycolor5]
		table[row sep=crcr]{%
			10	16.1616145307965\\
			20	12.7620775978667\\
			30	10.8316150101077\\
			40	7.99293840595634\\
			50	6.39418461906147\\
			60	3.0713319102271\\
			70	2.66568413703025\\
			80	2.56831411762501\\
		};
		
		\addplot [color=mycolor6]
		table[row sep=crcr]{%
			10	15.354265723374\\
			20	12.538895159837\\
			30	11.2617454229969\\
			40	8.44647269574381\\
			50	7.55886226515842\\
			60	5.92950461657826\\
			70	4.75469897329019\\
			80	3.52686733075153\\
		};
		
	\end{axis}

	\begin{axis}[%
		width=2.80in,
		height=2.40in,
		at={(3.5in,0.469in)},
		scale only axis,
		xmin=10,
		xmax=80,
		xlabel style={font=\color{mycolor7}},
		xlabel={gap length (ms)},
		ymin=-3,
		ymax=0,
        ytick distance=0.5,
		ylabel style={font=\color{mycolor7}},
		ylabel={ODG},
		axis background/.style={fill=white},
		title style={font=\bfseries\color{mycolor7}},
		title={PEMO-Q ODG},
		xmajorgrids,
		ymajorgrids,
		legend style={
            at={(0.5,-0.25)},
            anchor=north,
            legend cell align=left,
            align=left,
            legend columns=2,
            transpose legend,
            /tikz/every even column/.append style={column sep=2mm},
            /tikz/every odd column/.append style={column sep=1mm}}
		]
		\addplot [color=mycolor1]
		table[row sep=crcr]{%
			10	-0.0361542424770202\\
			20	-0.145977626458787\\
			30	-0.390725684445011\\
			40	-0.771197153542244\\
			50	-1.22011183053611\\
			60	-1.6518515585394\\
			70	-2.04997239270736\\
			80	-2.23033913504493\\
		};
		\addlegendentry{extrapolation-based, $K = 2048$, Burg algorithm}
		
		\addplot [color=mycolor2]
		table[row sep=crcr]{%
			10	-0.0310579713470681\\
			20	-0.115663148090292\\
			30	-0.299561607363113\\
			40	-0.611299948812533\\
			50	-1.01605132215585\\
			60	-1.45545127708884\\
			70	-1.8268929455971\\
			80	-2.03219458741717\\
		};
		\addlegendentry{Janssen, gap-wise, $K = 2048$, Burg algorithm}
		
		\addplot [color=mycolor3]
		table[row sep=crcr]{%
            10	-0.0459897278941861\\
            20	-0.220569489379194\\
            30	-0.516412656316135\\
            40	-0.941471216079882\\
            50	-1.88497940151254\\
            60	-2.4564845288054\\
            70	-2.42519769325549\\
            80	-2.62937470943985\\
		};
		\addlegendentry{Janssen, rect.\ window, $K = 512$, Burg algorithm}
		
		\addplot [color=mycolor4]
		table[row sep=crcr]{%
            10	-0.0378933983314482\\
            20	-0.165856752911012\\
            30	-0.481664713075842\\
            40	-0.947541867528085\\
            50	-1.47042581400567\\
            60	-1.93319652016359\\
            70	-2.20915983462033\\
            80	-2.47918935818701\\
		};
		\addlegendentry{Etter, $K = 2048$, Burg algorithm}
		
		\addplot [color=mycolor5]
		table[row sep=crcr]{%
			10	-0.0556506158616167\\
			20	-0.200682237145153\\
			30	-0.5412033818748\\
			40	-1.25514473787089\\
			50	-1.76678641952239\\
			60	-2.34903899445193\\
			70	-2.69062061608635\\
			80	-2.82875841697913\\
		};
		\addlegendentry{A-SPAIN}
		
		\addplot [color=mycolor6]
		table[row sep=crcr]{%
			10	-0.0979117144940085\\
			20	-0.33091914009669\\
			30	-0.772592112927194\\
			40	-1.43057298074778\\
			50	-1.78816550587453\\
			60	-2.17213020106928\\
			70	-2.61035578917115\\
			80	-2.78865155975933\\
		};
		\addlegendentry{A-SPAIN-MOD}
		
	\end{axis}

	\begin{axis}[%
		width=2.80in,
		height=2.40in,
		at={(7in,0.469in)},
		scale only axis,
		xmin=10,
		xmax=80,
		xlabel style={font=\color{mycolor7}},
		xlabel={gap length (ms)},
		ymin=-1.5,
		ymax=0,
		ylabel style={font=\color{mycolor7}},
		ylabel={ODG},
		axis background/.style={fill=white},
		title style={font=\bfseries\color{mycolor7}},
		title={PEAQ ODG},
		xmajorgrids,
		ymajorgrids,
		legend style={at={(0.03,0.03)}, anchor=south west, legend cell align=left, align=left}
		]
		\addplot [color=mycolor1]
		table[row sep=crcr]{%
			10	-0.145262700726528\\
			20	-0.302477787489241\\
			30	-0.410315400663417\\
			40	-0.489093762615499\\
			50	-0.551898331894106\\
			60	-0.591739735448844\\
			70	-0.669891266341316\\
			80	-0.756772881949664\\
		};
		
		\addplot [color=mycolor2]
		table[row sep=crcr]{%
			10	-0.159643885747593\\
			20	-0.319767668370557\\
			30	-0.425130606554472\\
			40	-0.480880618167751\\
			50	-0.518622055356956\\
			60	-0.555544728254786\\
			70	-0.630799509049537\\
			80	-0.696379179147104\\
		};
		
		\addplot [color=mycolor3]
		table[row sep=crcr]{%
            10	-0.12753788855158\\
            20	-0.351849121096847\\
            30	-0.431169439407038\\
            40	-0.467091306061312\\
            50	-0.604932042771311\\
            60	-0.72720314209439\\
            70	-0.733268035512564\\
            80	-0.72776819635778\\
		};
		
		\addplot [color=mycolor4]
		table[row sep=crcr]{%
            10	-0.13280566891691\\
            20	-0.297070407392288\\
            30	-0.391503411959217\\
            40	-0.446344361231182\\
            50	-0.494290682162397\\
            60	-0.516955240019193\\
            70	-0.599228060563679\\
            80	-0.636870820411379\\
		};
		
		\addplot [color=mycolor5]
		table[row sep=crcr]{%
			10	-0.267977607378664\\
			20	-0.402634337340751\\
			30	-0.527681073495913\\
			40	-0.64117252046808\\
			50	-0.737253043553519\\
			60	-0.802344610448469\\
			70	-0.888939218613272\\
			80	-1.05674995440483\\
		};
		
		\addplot [color=mycolor6]
		table[row sep=crcr]{%
			10	-0.394495809459359\\
			20	-0.512614462927672\\
			30	-0.632752143674424\\
			40	-0.720677287182292\\
			50	-0.775313208692802\\
			60	-0.789975625920146\\
			70	-0.891854896702119\\
			80	-0.975288225378978\\
		};
		
	\end{axis}

\end{tikzpicture}%
}
        \caption{Testing Etter versus baselines. Note the cropped vertical axes in the case of PEMO-Q ODG and PEAQ ODG; the full scale ranges from $-4$ to $0$.}
        \label{fig:maintest_plot_with_spain_shorter}
    \end{figure*}
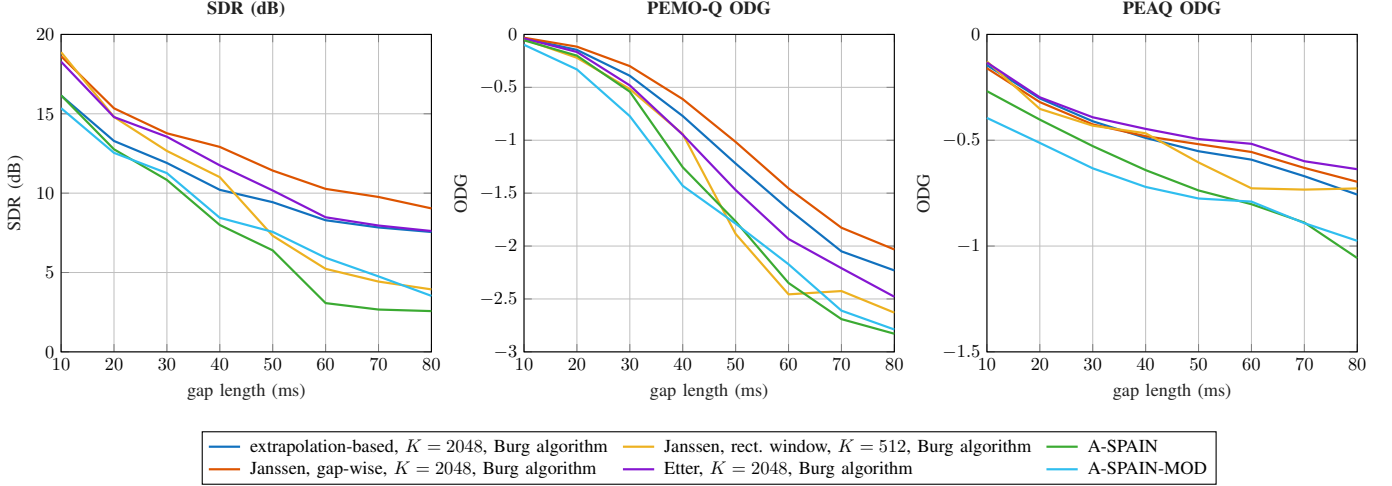

    Regarding computation time, Fig.~\ref{fig:maintest_plot_by_method_shorter_time} shows that both the model order and the gap length have an influence.
    While Burg is slower than LPC, for long gaps this difference becomes negligible (except for extrapolation), because the runtime is dominated by other algorithmic components; for Etter method in particular, the main cost is solving the linear system \eqref{eq:Dxy}, whose size and structure do not depend on the chosen AR model estimator.    
    
    For broader comparison with reference methods,
    we consider the windowed and gap-wise Janssen methods \cite{javevr86, MokryRajmic2025:Inpainting.AR} and sparsity-based A-SPAIN \cite{MokryZaviskaRajmicVesely2019:SPAIN} and A-SPAIN-MOD \cite{TaubockRajbamshiBalasz2021:SPAINMOD}.
    In this context,
    Etter ranks among the stronger approaches and is even the top-performing method according to PEAQ (see Fig.~\ref{fig:maintest_plot_with_spain_shorter}).
    
\begin{figure}[htb]
        \centering
        \adjustbox{width=\linewidth}{%
        \input{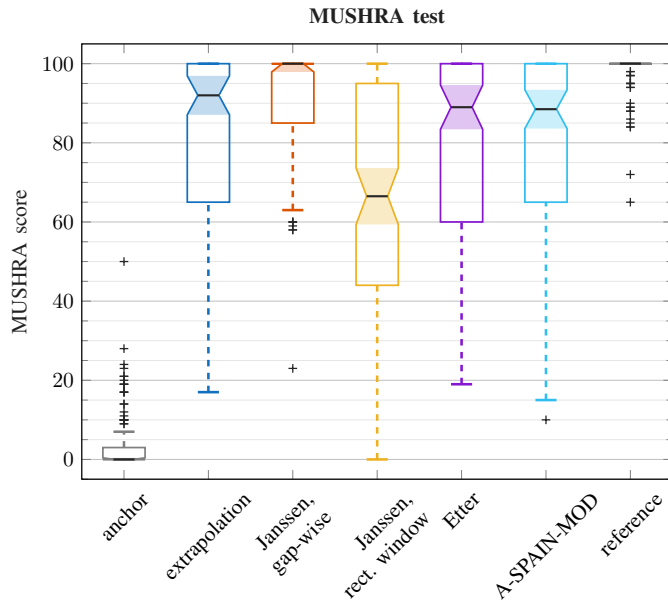}}
        \caption{A boxplot showing the distribution of scores in the MUSHRA listening test. The color coding is in correspondence with Fig.~\ref{fig:maintest_plot_with_spain_shorter}. For results per gap length, see the accompanying webpage.}
        \label{fig:mushra}
    \end{figure}

    To validate the objective results, we also conducted a~MUSHRA-like listening test \cite{ITU-R2015:MUSHRA} using the webMUSHRA environment \cite{Schoeffler2018:webMUSHRA} on a~subset of test signals (three signals: clarinet, violin, piano; three gap lengths: 20, 50, and 80~ms; the SPAIN method omitted).
    The test involved 19 assessors, pre-screened as a~knowledgeable public; ratings were collected under partially controlled conditions (participants used their own devices, but with a~quiet environment ensured and headphones required).
    In post-screening, 14 assessors passed by giving at least 7 out of 9 hidden reference signals a~score of at least 90.
    The resulting score distributions are shown in Fig.~\ref{fig:mushra}.
    Overall, the subjective scores mostly agree with the ranking implied by PEMO-Q ODG (see the middle plot of Fig.~\ref{fig:maintest_plot_with_spain_shorter}):
    gap-wise Janssen \cite{MokryRajmic2025:Inpainting.AR} is superior, followed by extrapolation and Etter with no significant difference.
    However, unlike the objective evaluation, the listening test favors A-SPAIN-MOD \cite{TaubockRajbamshiBalasz2021:SPAINMOD} more strongly, placing it on par with extrapolation and Etter.

	\section{Conclusion}
    \label{sec:conclusion}

    The paper revisits Etter method for audio signal inpainting based on the forward and backward AR prediction,
    which has been disregarded by recent literature on audio restoration.
    The method is briefly introduced, followed by two generalizations:
    The first aims at allowing the AR model order to exceed the length of the gap to be filled;
    this is crucial as recent works mostly prefer high-order models.
    The second generalization aims at the PLC use case,
    where causality and tight latency constraints need to be maintained.
    Reasonable derivation from Etter method is possible using the left-sided context and AR model parameters only.
    However, it can be shown that this modification results in a~straightforward approach of sample-wise signal extrapolation.

    Both objective and subjective results of an experiment on inpainting gaps in musical signals (up to 80\,ms) indicate that Etter method is competitive with other AR-based inpainting methods.
    Importantly, gap-wise Janssen method remains superior in terms of reconstruction quality,
    while extrapolation-based inpainting, scoring similar to Etter, is preferable in terms of computation time.
    As future work, the structure of the matrices used within the interpolation method promises numerically efficient solution to the linear system \eqref{eq:Dxy},
    bringing the computational load of Etter closer to extrapolation.
    
	


    \bibliographystyle{ieeetr}
    \bibliography{literatura}

\end{document}